\documentclass[print]{mn2e}
\usepackage{epsfig}
\usepackage{color}

\begin{document}
\title[The LF of {\it Swift} GRBs]{The luminosity function of {\it Swift} long gamma-ray bursts}
\author[Cao et al.]{Xiao-Feng Cao$^{1}$, Yun-Wei Yu$^{1,2}$\thanks{yuyw@phy.ccnu.edu.cn}, K. S. Cheng$^{2}$, and Xiao-Ping Zheng$^{1}$
\\$^1$Institute of Astrophysics, Central China Normal
University, Wuhan, 430079, China
\\$^2$Department of Physics, The University of Hong Kong,
Pokfulam Road, Hong Kong, China} \maketitle

\begin{abstract}
The accumulation of {\it Swift} observed gamma-ray bursts (GRBs)
gradually makes it possible to directly derive a GRB luminosity
function (LF) from observational luminosity distribution, where
however two complexities must be involved as (i) the evolving
connection between GRB rate and cosmic star formation rate and (ii)
observational selection effects due to telescope thresholds and
redshift measurements. With a phenomenological investigation on
these two complexities, we constrain and discriminate two popular
competitive LF models (i.e., broke-power-law LF and single-power-law
LF with an exponential cutoff at low luminosities). As a result, we
find that the broken-power-law LF could be more favored by the
observation, with a break luminosity $L_b=2.5\times10^{52}\rm
erg~s^{-1}$ and prior- and post-break indices $\nu_1=1.72$ and
$\nu_2=1.98$. For an extra evolution effect expressed by a factor
$(1+z)^{\delta}$, if the matallicity of GRB progenitors is lower
than $\sim0.1Z_{\odot}$ as expected by some collapsar models, then
there may be no extra evolution effect other than the metallicity
evolution (i.e., $\delta$ approaches to be zero). Alternatively, if
we remove the theoretical metallicity requirement, then a
relationship between the degenerate parameters $\delta$ and
$Z_{\max}$ can be found, very roughly,
$\delta\sim2.4(Z_{\max}/Z_{\odot}-0.06)$. This indicates that an
extra evolution could become necessary for relatively high
metallicities.
\end{abstract}

\section{Introduction}
Some confirmed associations between gamma-ray
bursts\footnote{Throughout we refer only ``long" gamma-ray bursts
with $T_{90}
> 2$ s, where $T_{90}$ is the interval observed to contain 90\% of the prompt emission.}
(GRBs) and Type Ib/c supernovae (Stanek et al. 2003; Hjorth et al.
2003; Chornock et al 2010) robustly suggest that GRBs are powered by
the collapse of the core of massive stars, which is also widely
accepted in theory (Woosley 1993; Paczy\'{n}ski 1998; Fryer et al.
1999; Wheeler et al 2000; Woosley \& Bloom 2006). In other words,
the detection of each GRB provides a witness of the death of a
massive star. Moreover, the intense brightness of GRBs makes them
detectable even at the edge of the universe (the highest redshift of
GRBs is $\sim9.4$ as reported by Cucchiara et al. 2011). So GRBs can
in principle be used as a tracer of the cosmic star formation
history. The crucial problem is whether GRBs are an unbiased tracer
or, more directly, how to calibrate GRB event rate to star formation
rate (SFR). On one hand, the cosmic evolution of metallicity could
be involved. This is because a very high angular momentum is
required for GRB progenitors and meanwhile massive stars in
lower-metallicity environments are less likely to loss much angular
momentum through stellar winds (e.g., Meynet et al. 1994; Langer \&
Henkel 1995; Vink \& de Koter 2005; MacFadyen \& Woosley 1999;
Wooseley \& Heger 2006). On the other hand, the luminosity function
(LF) of GRBs can also play an important role in the conversion from
the observed GRB redshift distribution to GRB formation history,
since the luminosity selection by telescopes leads to a lower
detection probability for higher-redshift GRBs.

To directly derive a GRB LF was impossible before the launch of {\it
Swift} (Gehrels et al. 2004), since there were only quite a few GRBs
whose redshifts are measured. A possible alternative way invokes
some luminosity-indicator relationships to avoid redshift
measurements (e.g. Yonetoku et al. 2004; Fenimore \& Ramirez-Ruiz
2000; Firmani et al. 2004), but the robustness of those indicators
may be not high enough. A much more popular method is to assume a LF
form with a few model parameters and then to fit the flux
distribution of the observed GRBs ($\log N-\log P$ distribution;
Schmidt 1999; Porciani \& Madau 2001; Firmani et al. 2004; Guetta et
al. 2005; Natarajan et al. 2005; Daigne et al. 2006; Salvaterra \&
Chincarini 2007; Salvaterra et al. 2009; Campisi et al. 2010).
Correspondingly, before {\it Swift}, a very large sample of GRBs had
been provided by the Burst and Transient Source Experiment (BATSE)
on board {\it Compton Observatory}. Nevertheless, by such a fitting
to $\log N-\log P$ distribution only, it is not easy to eliminate
the degeneracy among the model parameters and even to determine the
form of the LF. As a result, two competitive LF models as a
broken-power law (BPL) and a single-power law with an exponential
cutoff (SPLEC) at low luminosities are usually adopted in
literature.

Thanks to {\it Swift} spacecraft, in the past few years the number
of GRBs with measured redshift grows rapidly. This makes it possible
to provide more stringent constraints on the LF parameters (Daigne
et al. 2006; Salvaterra \& Chincarini 2007; Salvaterra et al. 2009;
Campisi et al. 2010). The new constraints robustly rule out the
models in which GRBs unbiased trace the cosmic star formation or
GRBs are characterized by a constant LF. In other words, an
evolution effect is suggested. In view of the not-small size of the
{\it Swift} GRB sample, it has become possible to derive a GRB LF
only with {\it Swift} GRBs. Very recently, Wanderman \& Piran (2010)
tried to directly convert the luminosity distribution of {\it Swift}
GRBs to a LF, without a prior assumed LF form and without a help
from the BATSE data
In such a LF-determination process, the treatments on
observational selection effects play a very important role.
Meanwhile, a possible extra evolution effect should also be paid
much attention to. In this paper, with a phenomenological
investigation on the evolution effect and the selection effects, we
constrain and discriminate the BPL and SPLEC models by using {\it
Swift} observed GRBs.

In the next section, some observational and theoretical materials
related to the GRB LF are described. In section 3, the evolution
effect is constrained and analyzed with relatively-high-luminosity
GRBs. In Section 4, firstly, we derive an initial LF in both the BPL
and SPLEC models by directly fitting the observational luminosity
distribution of GRBs. Secondly, we analyze and constrain the
so-called redshift-desert effect with the initial LFs. Finally, the
combination of the above two processes gives a final constraint on
the GRB LF, with which the prior selected luminosity threshold is
checked. In Section 5, conclusion and discussion are given.

\section{Basic materials}
\subsection{{\it Swift} observed GRBs}
\begin{figure}
\resizebox{\hsize}{!}{\includegraphics{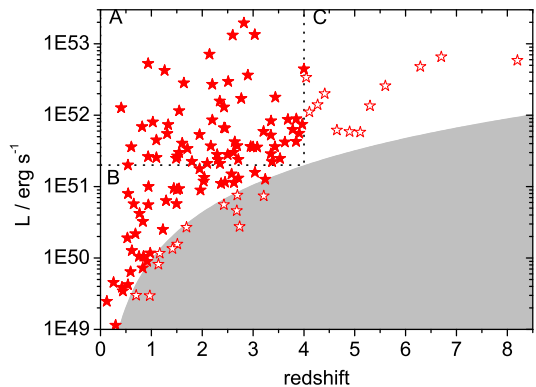}}
\caption{Luminosity-redshift distribution of 125 {\it Swift} GRBs
with redshift, where the shaded region represents the luminosity
threshold adopted in our calculations (see Equation \ref{Lth} and
explanations there). The 24 data labeled by open asterisks will be
excluded from our most statistics.}
\end{figure}
\begin{figure}
\resizebox{\hsize}{!}{\includegraphics{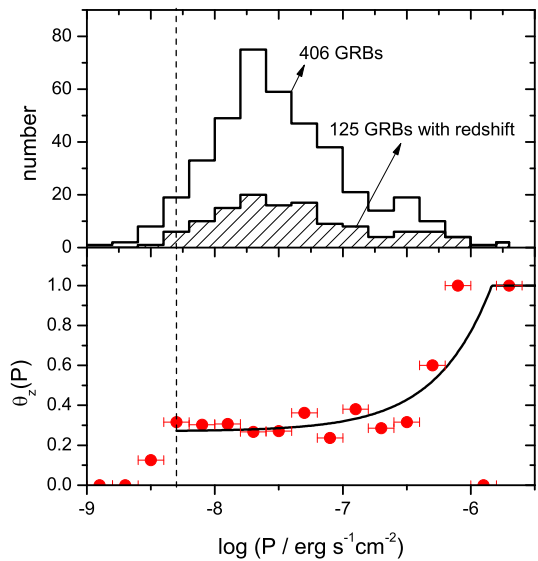}} \caption{{\it
Upper panel}: Comparison between the distributions of average fluxes
of the total 406 GRBs and the 125 GRBs with redshift; {\it Lower
panel}: Probability of redshift measurements as a function of flux
and an empirical fitting given by Equation (\ref{redshiftpro}),
where the horizon error bars correspond to the bin width. The
vertical dashed line represents the selected threshold $P _{\rm
th}=5\times10^{-9}~\rm erg~s^{-1}cm^{-2}$.}
\end{figure}
In the past few years, {\it Swift} has greatly promoted our
understanding of GRBs. Here we take GRBs with measured redshift $z$
from the {\it Swift}
archive\footnote{http://swift.gsfc.nasa.gov/docs/swift/archive/grb\_table.}.
For most of these GRBs till GRB 090813, their spectral peak energy
$E_p$ and isotropically-equivalent energy release $E_{\rm iso}$ in
the burst rest-frame $\rm 1-10^4~keV$ band have been provided by
Butler et al. (2007, 2010). However, it should be noticed that, due
to the narrow energy bandpass of the {\it Swift} Burst Alert
Telescope (BAT), the burst spectral parameters in Butler et al.
(2007, 2010) are actually estimated by the Bayesian statistics, but
not directly observed. Anyway, as in Kistler et al. (2008, 2009) and
Wang \& Dai (2009), an {\it average} luminosity can be roughly
estimated for these GRBs by
\begin{eqnarray}
L_{[1-10^{4}\rm keV]}={E_{\rm iso[1-10^{4}\rm keV]}\over
T_{90}/(1+z)}.\label{estL}
\end{eqnarray}
In our statistics, GRBs with $L<10^{49}\rm erg~s^{-1}$ will be
excluded, since they may belong to a distinct population called
low-luminosity GRBs (Soderberg et al. 2004; Cobb et al. 2006;
Chapman et al. 2007; Liang et al. 2007). Finally, 125 GRBs are
selected and their luminosity-redshift distribution is shown in
Figure 1. Correspondingly, including the GRBs without redshift,
there are totally 406 GRBs detected by {\it Swift} till GRB 090813.

In the upper panel of figure 2, we present the distributions of the
observed average fluxes\footnote{The average fluxes are calculated
from $S_{[15-150\rm keV]}/T_{90}$ where $S_{[15-150\rm keV]}$ is the
observed fluences.}, $P _{[15-150\rm keV]}$, for both the total 406
GRBs and the 125 GRBs with redshift. The ratio between these two
distributions generally displays the flux-dependence of the GRB
redshift measurements and, as shown in the lower panel, such a
redshift detection probability can be empirically expressed by
\begin{eqnarray}
\vartheta_{z}(P)=\min\left[0.27+{P \over 2.0\times10^{-6} \rm
erg~s^{-1}cm^{-2}},1\right].\label{redshiftpro}
\end{eqnarray}
A similar result has also been given in Qin et al. (2010). On the
other hand, the redshift detection probability may also depend on
redshift itself, which will be investigated in section 4.2.

\subsection{Model}
The luminosity-redshift distribution of GRBs is determined by both
the LF $\phi(L)$ and the comoving rate $\dot{R}(z)$ of GRBs, which
are respectively defined by
\begin{eqnarray}
\frac{d{N}}{dL}=\phi(L),
\end{eqnarray}
and
\begin{eqnarray}
\frac{d\dot{N}}{dz}=\dot{R}(z)\frac{dV(z)/dz}{1+z},\label{}
\end{eqnarray}
where the dot represents time derivation, the factor $(1+z)$ is due
to the cosmological time dilation of the observed rate and
$dV(z)/dz$ is the comoving volume element. In the standard
$\Lambda$-cold dark matter cosmology, $dV(z)/dz=4\pi d_c(z)^2
c/H(z)$ with $d_c(z)=d_l(z)/(1+z)$, where the luminosity distance
reads $d_l(z)=c(1+z)\int_0^{z}H(z')^{-1}dz'$ with
$H(z')=H_0[(1+z')^3\Omega_{m,0}+\Omega_{\Lambda,0}]^{1/2}$.
Throughout we adopt the cosmological parameters as
$\Omega_{m,0}=0.27$, $\Omega_{\Lambda,0}=0.73$, and $H_0=71~\rm
km~s^{-1}Mpc^{-1}$ (Komatsu et al. 2010).

Firstly, for the GRB rate $\dot{R}(z)$, it can in principle be
connected to the cosmic SFR $\dot{\rho}_*(z)$, since in the
collapsar model the formation of each GRB just indicates the death
of a short-lived massive star. For relatively low redshifts ($z<
4$), the SFR can be expressed approximately by (Hopkins \& Beacom
2006)
\begin{equation}
\dot{\rho}_*(z)\propto\left\{
\begin{array}{ll}
(1+z)^{3.44},&z<0.97,\\
(1+z)^{-0.26},&0.97\leq z< 4,
\end{array}\right.\label{sfh1}
\end{equation}
with $\dot{\rho}_*(0)=0.02~{\rm M_{\odot}yr^{-1}Mpc^{-3}}$, whereas
the star formation history above $z\sim4$ is unclear so far. So 12
GRBs with $z>4$ (the data in region C in Figure 1) are excluded from
our statistics. Secondly, for the GRB LF, two representative forms
are usually assumed in literature as a BPL
\begin{equation}
\phi(L)\propto\left\{
\begin{array}{ll}
\left({L\over L_b}\right)^{-\nu_1},&L\leq L_b,\\
\left({L\over L_b}\right)^{-\nu_2},&L>L_b.
\end{array}\right.,\label{bpl}
\end{equation}
and a SPLEC
\begin{equation}
\phi(L)\propto \left({L\over
L_{p}}\right)^{\nu}e^{-L_p/L}.\label{spl}
\end{equation}
Then the expected number of GRBs with redshift $z_1<z<z_2$ and
luminosity $L_1<L<L_2$ can be calculated by
\begin{eqnarray}
N^{\rm exp}\propto\int_{z_1}^{z_2}\int_{\max[L_1,L_{\rm
th}(z)]}^{L_2} (1+z)^\Delta\phi(L)\dot{\rho}_*(z)dL{dV(z)\over
1+z},\label{GRBnum_theory}
\end{eqnarray}
where the extra evolving factor $(1+z)^\Delta$ is introduced by
considering that (i) the connection between the GRB rate and the SFR
could be not in a trivial way and (ii) the LF could evolve with
redshift. Corresponding to different selection criterions and bin
methods for different GRB samples, the specific form of the above
equation (e.g., the sequence and the range of the integrals) should
be changed, see Equations (10), (11), (13), (15), and (18).

The luminosity threshold invoked in Equation (\ref{GRBnum_theory})
can be given by
\begin{eqnarray}
L_{\rm th}(z)=4\pi d_l(z)^2P _{\rm th}k(z),\label{Lth}
\end{eqnarray}
where $k(z)\equiv \int_{1\rm keV}^{10^{4}\rm
keV}S(E')E'dE'/\int_{15(1+z)\rm keV}^{150(1+z)\rm keV}S(E')E'dE'$
(the primes represent rest-frame energy) converts the observed flux
in the BAT energy band $\rm 15-150$ keV into the bolometric flux in
the rest-frame $\rm 1-10^4$ keV. The observed photon number spectrum
$S(E)$ can be well expressed by the empirical Band function (Band et
al. 1993), more simply, a broken power law. The value of $k$ varies
from 5.4 to 2.1 as the redshift increases from 0 to 10, by taking
the rest-frame peak energy as $E'_p\sim200~{\rm keV}$ (the most
frequent value in the Butler et al.'s database) and the photon
indices prior and post the break energy as 1 and 2.25 , respectively
(Preece et al. 2000). On the other hand, unfortunately, a precise
description for $P_{\rm th}$ is nearly impossible, since the trigger
of the BAT is very complicated and, especially for GRBs with
redshift, the actual threshold is determined by the combination of
the BAT and other related telescopes. Instead of an abrupt cutoff at
$P_{\rm th}$, a realistic situation could be that the detection
efficiency starts to remarkably decrease at a certain flux and
approaches to be zero with decreasing fluxes. Therefore, in the
following calculations, a relatively high value for $P _{\rm th}$ is
taken as $5\times10^{-9}~\rm erg~s^{-1}cm^{-2}$ and,
correspondingly, 12 GRBs below the selected threshold are further
excluded (as shown in Figure 1). Strictly speaking, the selected
$P_{\rm th}$ is not but higher than the true sensitivity. This can
make us avoiding the complex arising from the trigger probability.
The availability of the selected $P_{\rm th}$ will be checked by a
fitting to the $\log N-\log P$ distribution in section 4.3.
\subsection{Observational luminosity distribution}
\begin{figure}
\resizebox{\hsize}{!}{\includegraphics{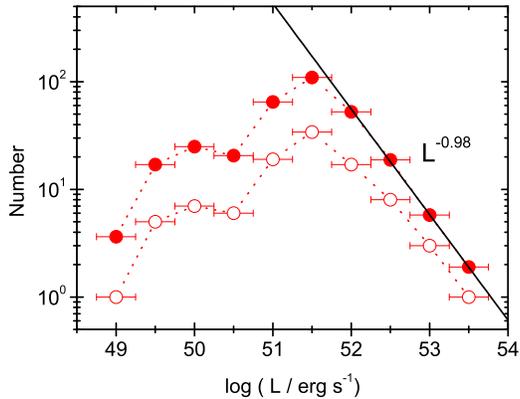}}
\caption{Luminosity distribution of the selected 101 GRBs, where the
horizon error bars correspond to the bin width. Open and solid
circles correspond to pre- and post-corrected distributions,
respectively, and the solid line gives a power-law fitting for the
corrected high-luminosity distribution. The equivalent number of the
corrected sample is about 319.}
\end{figure}
As the main objective we want to account for, the luminosity
distribution of the selected 101 GRBs (solid asterisks in Figure 1)
is presented in Figure 3, where an obvious break appears at
$\sim3\times10^{51}\rm erg~s^{-1}$. Such a break may reflect an
intrinsic break in LF or just be caused by the selection effects
arising from the BAT and also other related telescopes, which is
what we want to clarify in this paper. In order to avoid the
consideration of the flux-dependence of the redshift measurements in
our analyses, we count an effective number as $\vartheta_{z}^{-1}(P
)$ for each GRB with flux $P$ whose redshift is measured. As a
result, an effective GRB sample of a number of about 319 is derived
from the 101 GRBs. Since darker GRBs have higher weight in the
statistics, the corrected luminosity distribution becomes steeper,
especially above the break luminosity. A good power-law fitting as
$N\propto L^{-0.98}$ to the high-luminosity distribution indicates
$\nu_2=1.98$ in the BPL model and $\nu=1.98$ in the SPLEC model, in
view of the probable unimportance of most observational selection
effects at high-luminosity range. In the following Figures 4, 6, 7,
and 8, the observational data are all with the same number
correction.

\begin{figure}
\resizebox{\hsize}{!}{\includegraphics{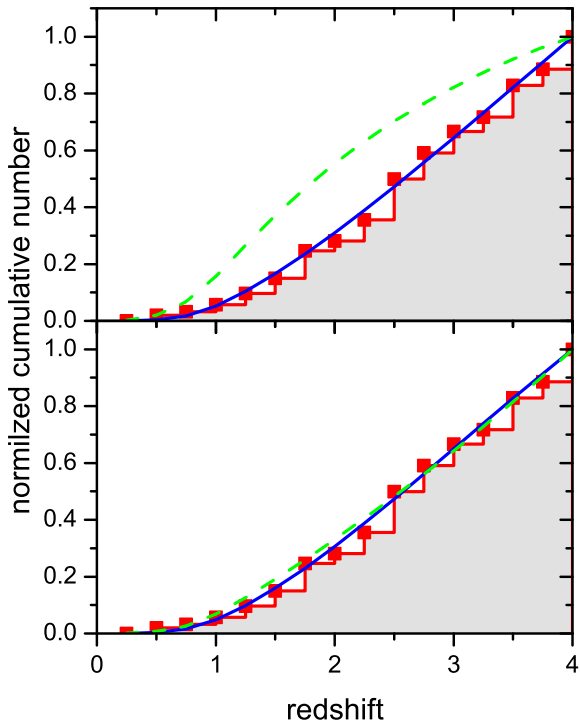}}
\caption{Normalized cumulative number of the 63 GRBs with $z<4$ and
$L>2\times10^{51}\rm erg~s^{-1}$ as a function of redshift
(histogram). {\it Upper panel}: Fittings to the redshift
distribution with Equation (\ref{fit_zdis_L51}), where the solid and
dashed lines correspond to $\Delta=1.93$ and $\Delta=0$,
respectively. {\it Lower panel}: Fittings to the redshift
distribution with Equation (\ref{fit_zdis_L51_2}), where the solid
and dashed lines correspond to $Z_{\max}=0.72Z_{\odot}$,
$\delta=1.56$ (best fit) and $Z_{\max}=0.1Z_{\odot}$, $\delta=0$,
respectively.}
\end{figure}
\section{Evolution effect}
Before we use the observational luminosity distribution to constrain
the GRB LF with equation (\ref{GRBnum_theory}), we should determine
the evolution parameter $\Delta$ in advance. Following Y\"uksel et
al. (2008) and Kistler et al. (2009), the value of $\Delta$ can be
constrained by fitting the observational cumulative redshift
distribution of GRBs with relatively high luminosities ($L\geq
L_{\rm cut}$). The cut luminosity $L_{\rm cut}$ is chosen to be
equal to or higher than the threshold at the highest redshift of the
sample (here $z_{\max}=4$), so that, in the corresponding
theoretical calculation, the integral of the LF can be treated as a
constant coefficient no matter the specific form of the LF, i.e.,
\begin{eqnarray}
N^{\rm exp}_{<z} \propto\int_{L_{\rm
cut}}^{L_{\max}}\phi(L)dL\int_0^z(1+z')^{\Delta}\dot{\rho}_*(z'){dV(z')\over1+z'}.\label{fit_zdis_L51}
\end{eqnarray}
Due to the limited size of the sample, the observational redshift
distribution actually is slightly dependent on the selected $L_{\rm
cut}$. So we take $L_{\rm cut}=L_{\rm th}(4)=2\times10^{51}\rm
erg~s^{-1}$ to reduce the statistical uncertainty as much as
possible. Consequently, 63 GRBs (the data in region A in Figure 1)
are selected. A comparison between the model and the observation is
presented in the upper panel of Figure 4, which shows that the
non-evolution case ($\Delta=0$) can be definitely ruled out, as
found before (e.g., Salvaterra \& Chincarini 2007; Salvaterra et al.
2009; Kistler et al. 2008, 2009). The best fitting to the
observation gives $\Delta=1.93$\footnote{If we do not correct the
GRB number by the factor $\vartheta_{z}^{-1}(P)$, we can get
$\Delta=1.44$, which is consistent with the results in Kistler et
al. (2008, 2009).}.

\begin{figure}
\resizebox{\hsize}{!}{\includegraphics{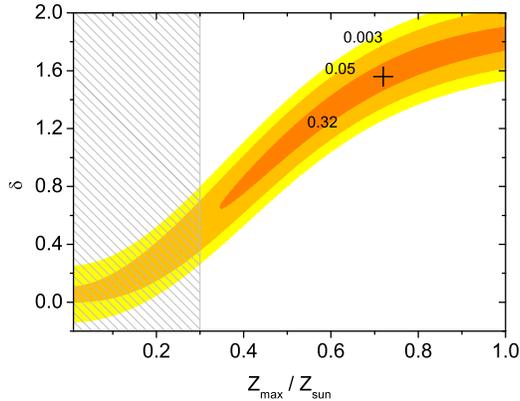}}
\caption{$\chi^2$-probability distribution of the fittings with
Equation (\ref{fit_zdis_L51_2}) to the redshift distribution of the
63 GRBs with $z<4$ and $L>2\times10^{51}\rm erg~s^{-1}$. The
best-fit parameters $\delta=1.56$ and $Z_{\max}=0.72Z_{\odot}$ is
labeled by the cross. The hatched region represents the theoretical
metallicity range expected by some collapsar models.}
\end{figure}
Then an interesting question arises as where such an evolution comes
from. As found by MacFadyen \& Woosley (1999) and Yoon et al.
(2006), the formation of the black hole (or neutron star) during the
collapse can drive a GRB event only if the collapsar has high
angular momentum. In order to avoid strong stellar winds loosing
angular momentum, GRB progenitors are required to be in
low-metallicity evironments (Woosley \& Heger 2006; Yoon et al.
2006). This theoretical metallicity requirement is widely favored by
the estimations of the metallicities of long GRB hosts (e.g. Chen et
al. 2005; Gorosabel et al. 2005; Starling et al. 2005). Therefore,
it is suggested that the observationally required evolution could be
mainly due to the cosmic evolution of metallicity. Specifically, as
derived by Langer \& Norman (2006), the fraction belonging to
metallicity below $Z_{\max}$ can be calculated by $\Psi_{<Z_{\rm
\max}}(z)={\left.\hat{\Gamma}[0.84,(Z_{\rm
\max}/Z_{\odot})^210^{0.3z}]\right./\Gamma(0.84)}$, where $Z_{\max}$
is the maximum metallicity available for GRB progenitors and
$\hat{\Gamma}$ and $\Gamma$ are the upper incomplete and complete
gamma functions. Following this consideration, Equation
(\ref{fit_zdis_L51}) becomes
\begin{eqnarray}
N^{\rm exp}_{<z} \propto\int_{L_{\rm
cut}}^{L_{\max}}\phi(L)dL\int_0^z(1+z')^{\delta}\Psi_{<Z_{\rm
\max}}(z')\dot{\rho}_*(z'){dV(z')\over1+z'}.\label{fit_zdis_L51_2}
\end{eqnarray}
Varying the parameters $Z_{\max}$ and $\delta$, we refit the
observational redshift distribution shown in Figure 4 and present
the distribution of the $\chi^2$-probabilities of the fittings in
Figure 5. At first sight, the best-fit parameters $Z_{\rm
\max}=0.72Z_{\odot}$ and $\delta=1.56$ may indicate that there is a
significant extra evolution other than the metallicity evolution.
However, the long and narrow contours shown in Figure 5 robustly
demonstrate that the parameters $Z_{\max}$ and $\delta$ are actually
strongly degenerate and, moreover, the specific values of the
best-fit parameters are probably sensitive to the selection of the
observational sample. Therefore, instead of paying attention to the
best-fit parameters, we treat the relationship between the two
parameters exhibited by the contours as a more valuable result, very
roughly, $\delta\sim2.4(Z_{\max}/Z_{\odot}-0.06)$. Anyway, an
independent constraint on these two parameters is demanded in order
to reduce the parameter degeneracy.

For example, a theoretical constraint on metallicity can be invoked
(e.g., Campisi et al. 2010). As proposed by Woosley \& Heger (2006)
and Yoon et al. (2006), the maximum metallicity available for GRB
progenitors is likely to be within $\sim[0.1-0.3]Z_{\odot}$. As
shown in Figure 5, for $Z_{\max}<0.3Z_{\odot}$, the value of
$\delta$ would not be higher than 0.8 with 99.7\% confidence.
Especially for $Z_{\max}<0.1Z_{\odot}$, the value of $\delta$
approaches to be zero. The fitting to the observation with
$Z_{\max}=0.1Z_{\odot}$ and $\delta=0$ is shown in the lower panel
of Figure 4 in comparison with the fitting with $Z_{\rm
\max}=0.72Z_{\odot}$ and $\delta=1.56$. As can be seen, the
difference between these two fittings is not very significant.
Therefore, an extra evolution other than the metallicity evolution
may be not inevitable if $Z_{\max}$ is indeed very low.

In the following calculations, we take the best-fit parameters
$Z_{\max}=0.72$ and $\delta=1.56$ just for a good description for
the evolution effect. Constraints on the LF actually can not be
significantly affected by the variation of $Z_{\max}$ and $\delta$
as long as they satisfy the required relationship. On the other
hand, for simplicity, we will ascribe the possible extra evolution
to some unknown factors in the connection between the GRB rate and
the SFR, i.e.,
\begin{eqnarray}
\dot{R}(z)=C_R(1+z)^\delta\Psi_{<Z_{\max}}(z)\dot{\rho}_*(z),
\end{eqnarray}
where the proportional coefficient $C_R$ will be determined in
Section 4.4. In other words, the LF will be taken to be non-evolving
in this paper.

\section{Luminosity function}
\subsection{A preliminary constraint}
With given $Z_{\max}$ and $\delta$, we can constrain the unknown LF
by fitting the observational luminosity distribution of the 101 GRBs
by
\begin{eqnarray}
N_{[L_1,L_2]}^{\rm
exp}\propto\int_{L_1}^{L_2}\int_0^{\min[z_{M}(L),4]}\phi(L)\dot{R}(z){dV(z)\over1+z}dL,\label{fit_lumdis}
\end{eqnarray}
which gives the expected number in each luminosity bin $L_1\leq
L\leq L_2$. The maximum redshift $z_M(L)$ as a function of
luminosity can be solved from\footnote{With an approximate
expression for luminosity distance as $d_l(z)\approx{3c\over
H_0}\sqrt{1+z}(\sqrt{1+z}-1)$, the maximum redshift can be
approximatively calculated by $z_{M}\approx{1\over
2}\left(\sqrt{1+4H_0d_{l,M}/3c}+2H_0d_{l,M}/3c-1\right)$.}
\begin{eqnarray}
d_{l,M}(L)&=&{c(1+z)\over
H_0}\int_0^{z_{M}}{1\over\sqrt{(1+z)^3\Omega_{m,0}+\Omega_{\Lambda,0}}}dz\nonumber\\
&=&\left({L\over4\pi P _{\rm th}k}\right)^{1/2}.
\end{eqnarray}
With fixed $\nu_2=1.98$ in the BPL model and $\nu=1.98$ in the SPLEC
model and minimizing the $\chi^2$-statistic of the fittings, we
obtain the best-fit parameters as $L_b=2.5\times10^{52}\rm
erg~s^{-1}$ and $\nu_1=1.67$ for the BPL model and
$L_p=2.5\times10^{49}\rm erg~s^{-1}$ for the SPLEC model. As shown
in Figure 6, the fitting with a BPL LF seems much better than the
one with a SPLEC LF. This impels us to favor the BPL model. However,
the apparent oscillation of the observational data, which can not be
explained by both models, still demands a much more elaborate
fitting.
\begin{figure}
\resizebox{\hsize}{!}{\includegraphics{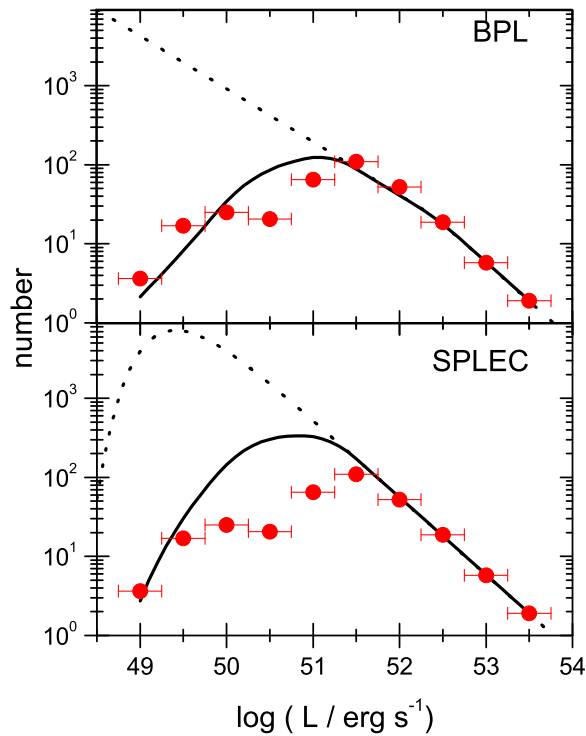}}\caption{The best
fitting to the observational GRB luminosity distribution with
equation (\ref{fit_lumdis}).
The dotted and solid lines represent intrinsic and observable
distributions, respectively. The parameters are
$L_b=2.5\times10^{52}\rm erg~s^{-1}$, $\nu_1=1.67$, and $\nu_2=1.98$
for the BPL model (upper) and $L_p=2.5\times10^{49}\rm erg~s^{-1}$
and $\nu=1.98$ for the SPLEC model (lower).}
\end{figure}

\subsection{Redshift-desert effect}
With the preliminary LFs derived above, we can give some
model-predicted cumulative redshift distributions by
\begin{eqnarray}
N^{\rm exp}_{<z}\propto \int_0^z\int_{L_{\rm
th}(z)}^{L_{\max}}\phi(L)\dot{R}(z')dL{dV(z')\over1+z'},\label{fit_zdis_Lall}
\end{eqnarray}
which are shown in the upper panel of Figure 7 in comparison with
the observational one of the 101 GRBs. Obviously, the observational
numbers at the middle redshifts are much less than the ones
predicted by both the models. Such a remarkable dip in the
observational redshift distribution is probably, at least partly,
related to the so-called `redshift-desert' effect, which is ignored
in the above analyses. As qualitatively analyzed by Fiore et al.
(2007), it could be difficult to measure redshifts within the range
$1.1<z<2.1$, since at $z>1.1$ some strong observable emission or
absorption lines are shifted outside the typical interval covered by
optical spectrometers ($3800-8000 {\AA}$) while Lyman-$\alpha$
enters the range at $z\sim2.1$. In this paper we do not try to give
a theoretical description for the redshift-desert effect, which must
involve many physical and technical issues. We also notice that a
same significant dip actually can not be found in the redshift
distribution of only GRBs with $L>2\times10^{52}\rm erg~s^{-1}$, as
shown in Figure 4. Hence, we suspect that the redshift-desert effect
may mainly influence the redshift measurements of relatively dark
GRBs, which can also be implied by the luminosity distribution.

\begin{figure}
\resizebox{\hsize}{!}{\includegraphics{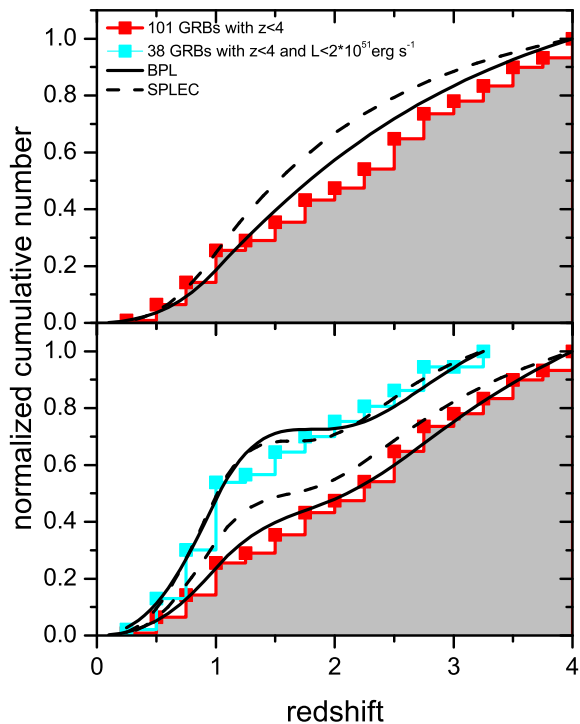}}\caption{Model-predicted
redshift distributions with (upper) and without (lower)
redshift-desert effect in comparison with the observational
distribution, where the adopted LFs are the same to Figure 6. The
parameters for the redshift-desert effect read $\mu=1.80$ and
$\sigma=0.79$ for the BPL model and $\mu=1.63$ and $\sigma=0.61$ for
the SPLEC model.}
\end{figure}
Therefore, we show the observational redshift distribution of only
38 GRBs with $z<4$ and $L<2\times10^{51}\rm erg~s^{-1}$ (the data in
region B in Figure 1) in the lower panel of Figure 7. Meanwhile, for
relatively dark GRBs, we tentatively suggest a Gaussian function
\begin{eqnarray}
\eta_{z}(z)=1-\exp\left[-{(z-\mu)^2\over\sigma^2}\right]\label{zdesert}
\end{eqnarray}
to phenomenologically describe the redshift-dependence of the
redshift measurements. In contrast, for sufficiently bright GRBs, we
take
\begin{eqnarray}
\eta_{z}(z)=1\label{zdesert2}.
\end{eqnarray}
Fittings to the distribution of the 38 GRBs give the best-fit
parameters as $\mu=1.80$ and $\sigma=0.79$ for the BPL model and
$\mu=1.63$ and $\sigma=0.61$ for the SPLEC model, which are
basically consistent with the theoretical expectation of the
redshift-desert effect.
With these phenomenological expressions of $\eta_{z}(z)$, we refit
the redshift distribution of the 101 GRBs, which is also shown in
the lower panel of Figure 7. As can be seen, the fittings are
greatly improved, as the observational dip in the redshift
distribution is produced naturally, especially in the BPL model.

\subsection{Final results}
\begin{figure}
\resizebox{\hsize}{!}{\includegraphics{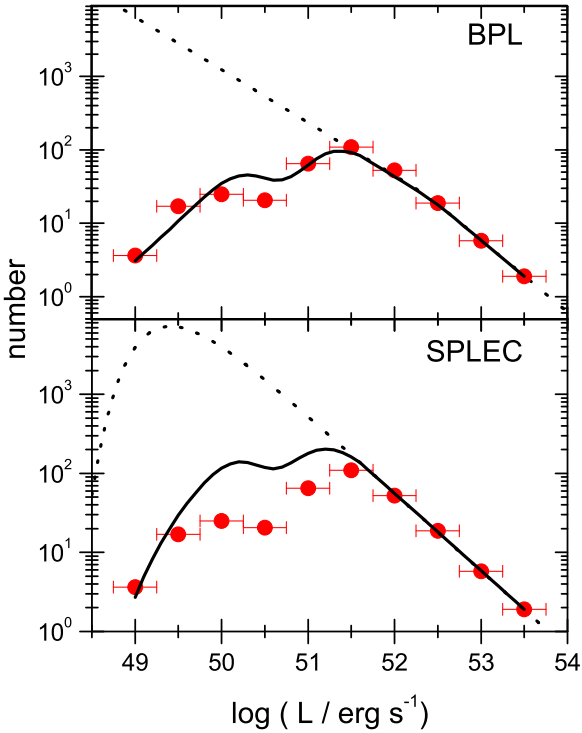}}\caption{The same
to Figure 6 but with a redshift-desert effect derived from Figure 7.
The LF parameters are $L_b=2.5\times10^{52}\rm erg~s^{-1}$,
$\nu_1=1.72$, and $\nu_2=1.98$ for the BPL model and
$L_p=2.5\times10^{49}\rm erg~s^{-1}$ and $\nu=1.98$ for the SPLEC
model.}
\end{figure}
Combining Equations (\ref{fit_lumdis}), (\ref{zdesert}), and
(\ref{zdesert2}), we refit the luminosity distribution of the 101
GRBs and find that $L_b=2.5\times10^{52}\rm erg~s^{-1}$ and
$\nu_1=1.72$ for the BPL model and $L_p=2.5\times10^{49}\rm
erg~s^{-1}$ for the SPLEC model. Strictly, we should use these new
LFs to re-constrain the redshift-desert effect and go recycling
until reaching a certain precision. For simplicity, we stop here
because the obtained new values of the parameters are only slightly
different from the preliminary ones. With these new parameters,
Figure 8 shows that the BPL model complies with the observation
successfully, whereas the SPLEC model still predicts some remarkable
excesses around $\sim10^{50}-10^{51}\rm erg~s^{-1}$. Therefore, we
prefer to conclude that the GRB LF could be a BPL.

Finally, with the derived BPL LF, we give a model-predicted
cumulative flux distribution by
\begin{eqnarray}
{N^{\rm
exp}_{>P}\propto\int_0^4\int_{L_P}^{L_{\max}}\phi(L)\dot{R}(z)\vartheta_{z}(P')\eta_{z}(z)dL{dV(z)\over1+z}},
\end{eqnarray}
where $L_{P}=4\pi d_l^2kP$, $P'=L/4\pi d_l^2k$, and the redshift
detection probability $\vartheta_z(P')\eta_z(z)$ has been determined
above. As shown in Figure 9, the consistency between the theoretical
and observational flux distributions indicates that our choice of
the luminosity threshold is basically reasonable, i.e., the trigger
probability above $P_{\rm th }$ by the BAT can be affirmed to nearly
constant.
\begin{figure}
\resizebox{\hsize}{!}{\includegraphics{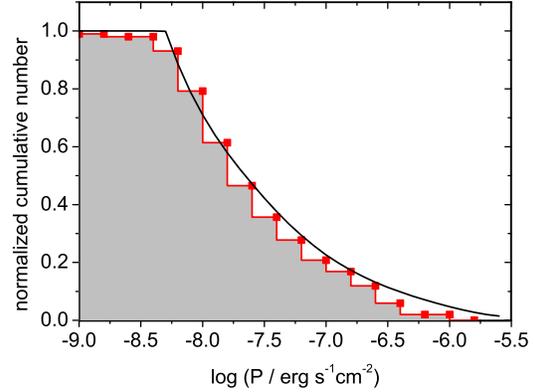}}\caption{Comparison
of the model-predicted flux distribution in the BPL LF model with
the observational one (without number correction) of the 101 GRBs.
The model parameters are the same to Figure 8.}
\end{figure}

\subsection{GRB rate}
In the above fittings to the luminosity distribution, we normalize
the model-predicted GRB number by the following
equation\footnote{The normalization is usually estimated with an
entire dataset or a good statistical point. Here, we select the data
at the highest luminosity $L=10^{53.5}\rm erg~s^{-1}$ for
normalization due to two reasons: (i) the data above $L=10^{52}\rm
erg~s^{-1}$ can be well fitted by a power law, which indicates that
all the data above $L=10^{52}\rm erg~s^{-1}$ are probably good
statistical, and (ii) more higher-luminosity GRBs may have less
selection effects.}:
\begin{eqnarray}
N_{[{53.25},{53.75}]}^{\rm
obs}&=&{\Delta\Omega\over4\pi}Tf_b\int_{{10^{53.25}}}^{10^{53.75}}C_L\phi(L)d
L
\int_0^{4}\dot{R}(z){dV(z)\over1+z}.\nonumber\\\label{normalization}
\end{eqnarray}
where $(\Delta\Omega/4\pi)\sim 0.1$ is the field of view of the BAT,
$T\sim5$ yr is the observational period, $f_b\sim0.01$ is the
beaming degree of the GRB outflow, and
$C_L^{}\approx{(\nu_1-1)}\left({L_{\min}\over
L_b}\right)^{\nu_1-1}{1\over L_b}$ is the normalization coefficient
of the BPL LF with $L_{\min}$ being an assumed minimum luminosity
for the GRBs.

For $\delta=1.56$, $Z_{\max}=0.72Z_{\odot}$, $\nu_1=1.72$,
$\nu_2=1.98$, $L_b=2.5\times10^{52}\rm erg~s^{-1}$, and
$N_{[{53.25},{53.75}]}^{\rm obs}=1.9$, the proportional coefficient
in the GRB rate can be constrained to
\begin{eqnarray}
C_R^{}=4\times10^{-6}\left({L_{\min}\over
10^{-4}L_b}\right)^{1-\nu_1}\left({f_b\over
0.01}\right)^{-1}M_{\odot}^{-1},
\end{eqnarray}
which yields a overall local GRB rate as
$\dot{R}(0)=C_R\Psi_{<Z_{\max}}(0)\dot{\rho}_*(0)=36(f_{b}/0.01)^{-1}~\rm
Gpc^{-3}yr^{-1}$ and an observed local GRB rate as
$f_b\dot{R}(0)=0.36~\rm Gpc^{-3}yr^{-1}$. This rate is basically
consistent with the previous results (e.g., Schmidt 2001; Guetta et
al. 2004, 2005; Liang et al. 2007; Wanderman \& Piran 2010).
The value of $C_R$ also implies that, besides the metallicity
requirement, GRB progenitors may also have some other
particularities. For example, as widely accepted, only massive
Wolf-Rayet stars (e.g., $> 20M_{\odot}$) are possible GRB
progenitors (MacFadyen \& Woosley 1999; Larsson et al. 2007). So a
small fraction arises as $f_{\rm
WR}=\left.\int_{20M_{\odot}}^{100M_{\odot}}\varphi(m)dm\right/\int_{0.1M_{\odot}}^{100M_{\odot}}m\varphi(m)dm\approx
2\times10^{-3}~M_{\odot}^{-1}$, where $\varphi(m)$ is the Salpeter
initial stellar mass function. Additionally, there is still an extra
factor
of $\sim10^{-3}-10^{-2}$ unexplained, which could be related to the
particularity of GRB progenitors in their rotation, magnetic fields,
etc.

\section{Conclusion and discussion}
The accumulation of {\it Swift} observed GRBs gradually makes it
possible to directly derive a GRB LF from observational luminosity
distribution, where however two complexities must be involved as (i)
the evolving connection between the GRB rate and the cosmic SFR and
(ii) observational selection effects. With a phenomenological
investigation on these two complexities, we constrain and
discriminate two popular competitive LF models and find that the BPL
LF model is more favored by the observation. However, in view of the
approximative description of the selection effects, the SPLEC still
can not be ruled out absolutely.

Although the derived values of the parameters $\mu$ and $\sigma$ are
basically consistent with the theoretical expectation of the
redshift-desert effect, the flux-dependence of the redshift-desert
effect is still very ambiguous (an abrupt luminosity boundary as
$2\times10^{51}\rm erg~s^{-1}$ is adopted in our analyses). More
seriously, the flux- and redshift-dependences of the redshift
measurements actually must be coupled with each other, but in our
analyses the functions $\vartheta_z(P)$ and $\eta_z(z)$ are
considered independently for simplicity. This may lead to an
overestimation of the redshift selection effect. Therefore, some
more elaborate theoretical considerations on the redshift
measurements are demanded. On the other hand, of course, a more
detailed analysis on the observational results would be helpful,
specially towards to every kind of redshift measurement methods.

Finally, our investigation on the evolution effect shows that, if
the matallicity of GRB progenitors is lower than $\sim0.1Z_{\odot}$
as expected by some collapsar models, there may be no extra
evolution effect (i.e., $\delta\sim0$) other than the metallicity
evolution. Alternatively, if we remove the theoretical metallicity
requirement,  a stronger extra evolution would be required for
higher metallicties. In the latter case, the extra evolution could
indicate some other evolutions in the GRB rate or indicate an
evolving LF which is not considered in this paper. To discriminate
these two possibilities is difficult but interesting. It will be
helpful to separate the GRB sample into few subsamples with
different redshift ranges and fit the luminosity distribution of
each subsamples one by one with an evolving LF. Such a further work
can be made as the GRB sample becomes large enough.
\section*{Acknowledgements}

The authors thank D. Yonetoku for his invaluable comments and
suggestions that have significantly improved our work. This work is
supported by the National Natural Science Foundation of China (grant
nos 11047121 and 11073008)
and by the Self-Determined Research Funds of CCNU (grant no.
CCNU09A01020) from the colleges' basic research and operation of MOE
of China. KSC is supported by the GRF Grants of the Government of
the Hong Kong SAR under HKU7011/09P.

\end{document}